\documentclass[conference]{IEEEtran}
\IEEEoverridecommandlockouts
\usepackage{amsmath,amssymb,amsfonts}
\usepackage{algorithmic}
\usepackage{graphicx}
\usepackage{textcomp}
\usepackage[dvipsnames]{xcolor}
\def\BibTeX{{\rm B\kern-.05em{\sc i\kern-.025em b}\kern-.08em
    T\kern-.1667em\lower.7ex\hbox{E}\kern-.125emX}}

\usepackage{comment}
\usepackage[printonlyused]{acronym}
\acrodef{MC}{molecular communication}
\acrodef{SF}{symmetry factor}
\acrodef{MDT}{magnetic drug targeting}
\acrodef{SPION}{superparamagnetic iron-oxide nanoparticle}
\acrodef{FWHM}{Full-Width-Half-Maximum}
\acrodef{VoFS}{Volume of Fluid Simulation}
\acrodef{LPS}{Lagrangian Particle Simulation}
\acrodef{ROI}{Region of Interest}

\usepackage{todonotes}
\usepackage[utf8]{inputenc}
\usepackage[normalem]{ulem} 

\usepackage{csquotes}

\usepackage[backend=biber,style=ieee,doi=false,isbn=false,mincitenames=1,maxcitenames=2]{biblatex}
\DeclareFieldFormat{sentencecase}{#1} 
\DeclareFieldFormat{titlecase}{#1} 
\DefineBibliographyStrings{english}{
	andothers = et~al\adddot\addspace
}

\usepackage{siunitx}
\makeatletter
\newcommand{\linebreakand}{%
  \end{@IEEEauthorhalign}
  \hfill\mbox{}\par
  \mbox{}\hfill\begin{@IEEEauthorhalign}
}
\makeatother
\usepackage{eso-pic}
\newcommand{\copyrightnotice}{%
  \AddToShipoutPictureFG*{%
    \AtPageUpperLeft{%
      \put(\LenToUnit{0.5\paperwidth},\LenToUnit{-0.9cm}){%
        \makebox[0pt][c]{%
          \parbox[t]{0.85\paperwidth}{\footnotesize
            Personal use of this material is permitted. This work has been submitted to the IEEE
            for possible publication. Copyright may be transferred without notice, after which
            this version may no longer be accessible.}%
        }%
      }%
    }%
  }%
}

\begin{document}
\copyrightnotice 

\title{Influence of the branching geometry on SPION propagation: coupling of experiment and simulation\\
\thanks{Funded by Deutsche Forschungsgemeinschaft (German Research Foundation) under the grant number 509922606 - GRK 2950: Synthetic Molecular Communication Across Different Scales: From Theory to Experiments, sub-project P05 and P06}
}

\author{\IEEEauthorblockN{Daniel Fleischhauer\IEEEauthorrefmark{1},
Luiz C. P. Wille\IEEEauthorrefmark{2},
Stefan Lyer\IEEEauthorrefmark{3},
Jens Kirchner\IEEEauthorrefmark{2}\textsuperscript{,}\IEEEauthorrefmark{4} and
Dietmar Drummer\IEEEauthorrefmark{1}}
\IEEEauthorblockA{\IEEEauthorrefmark{1}\textit{Institute of Polymer Technology (LKT)}, \IEEEauthorrefmark{2}\textit{Institute for Smart Electronics and Systems (LITES)},\\
\textit{Friedrich-Alexander-Universität Erlangen-Nürnberg (FAU)}, Erlangen, Germany\\
\IEEEauthorrefmark{3}\textit{Section for Experimental Oncology and Nanomedicine (SEON), ENT Department, University Hospital Erlangen}, Germany\\
\IEEEauthorrefmark{4}\textit{Fachhochschule Dortmund -- University of Applied Sciences and Arts}, Dortmund, Germany\\
Email: \{daniel.fleischhauer, luiz.wille, stefan.lyer, dietmar.drummer\}@fau.de, jens.kirchner@fh-dortmund.de}
}

\maketitle

\begin{abstract}
Magnetic Drug Targeting (MDT) is an emerging therapeutic concept in cancer medicine, where superparamagnetic iron oxide nanoparticles (SPIONs) are used as carriers for medication. Their magnetic properties allow for the manipulation of particle transport within the vascular structure and thus targeted delivery to the tumor site. To develop effective concepts for particle steering and accumulation, a detailed understanding of the particle--flow interaction, especially at asymmetric vascular bifurcations, is needed. Based on hierarchical scaling laws, generalized bifurcation geometries with varying degrees of asymmetry were derived and additively manufactured. The particle distribution was investigated by in situ optical measurements under different flow conditions and magnetic field configurations and compared with Lagrangian particle and Volume of Fluid Simulations (VoFS). While the particle mass partition followed the hydraulic flow split in asymmetric bifurcations, density-driven stratification dominated in the symmetric bifurcation, where one daughter branch received only a minor fraction of the particle mass at low flow rates despite carrying half of the flow. The Lagrangian model reproduced the predominantly advective propagation, whereas only the VoFS approach captured the stratification-induced asymmetry. Magnetic actuation counteracted the stratification, reversed the passive branch preference and increased particle retention. From a molecular communication perspective, bifurcation geometry, collective volume effects and magnetic actuation thus jointly determine the received particle signal at vascular bifurcations.

\end{abstract}


\section{Introduction}
Magnetic drug targeting (MDT) aims for efficient localized drug delivery by using \acp{SPION} as magnetically steerable carriers. At the tumor site, \acp{SPION} are accumulated by the magnetic field and release the medication. On their way, particles have to pass successive vascular bifurcations where they are affected by local flow partitioning. From a \ac{MC} perspective, \acp{SPION} can be interpreted as information carriers propagating through a flow-driven vascular channel, where the received particle signal is governed by the underlying hydrodynamics, the injection process, particle-specific transport mechanisms, and external magnetic actuation. Experimental \ac{MC} testbeds have established the use of \acp{SPION} in controlled pipe-flow systems \cite{Wicke.2022}, while recent studies have investigated effects such as pulsatile flow \cite{Wille2025a}, injection design \cite{Wille2025}, and sedimentation \cite{Xiao2026a}. In parallel, physical vessel models have been developed to study particle transport in branched vascular structures, demonstrating that vessel geometry and bifurcation topology can substantially affect particle distribution \cite{Fleischhauer2026, Thomas.2014, Amili.2019}.
Accurate prediction of particle propagation therefore requires channel models that extend beyond straight vessels and idealized advection–diffusion transport. Existing approaches range from particle-based descriptions of nanoparticle transport in branched vessels \cite{Tan.2013} to experimentally validated MC models for complex vessel networks \cite{Jamali2019}, but are based on idealised assumptions. However, experimentally grounded descriptions of how branch asymmetry affects SPION propagation and branch-specific concentration remain limited. This is particularly relevant because vascular networks exhibit non-uniform branching across spatial scales \cite{Tekin2016}, while concentrated SPION suspensions may show sedimentation and stratification behavior related to collective volume effects \cite{Xiao2026a, Fleischhauer2026}. Consequently, the relation between local hydrodynamic flow partitioning, particle redistribution, and magnetic steering at individual bifurcations remains insufficiently understood.
In this work, we investigate SPION propagation through generalized single-bifurcation vessel models with systematically varied asymmetry. Based on previously introduced generalized vascular geometries \cite{Fleischhauer2026}, additively manufactured flow models are combined with in situ optical measurements under different inlet flow rates and magnetic-field conditions. The experimental results are compared with both a Lagrangian particle simulation and an Eulerian \acf{VoFS} approach to distinguish predominantly advective particle transport from concentration-field effects such as stratification and sedimentation. By evaluating propagation time, temporal dispersion, branch-specific mass fraction, and post-peak washout, the study provides an experimentally grounded assessment of how branching asymmetry and stratification determine the passive particle routing, which of these mechanisms are captured by either simulation models and to which extent a magnetic field can reverse the passive routing towards the hydrodynamically disfavored branch.

\section{Geometry Design and Experimental Setup}
\begin{figure}
    \centering
    \includegraphics[width=1\linewidth]{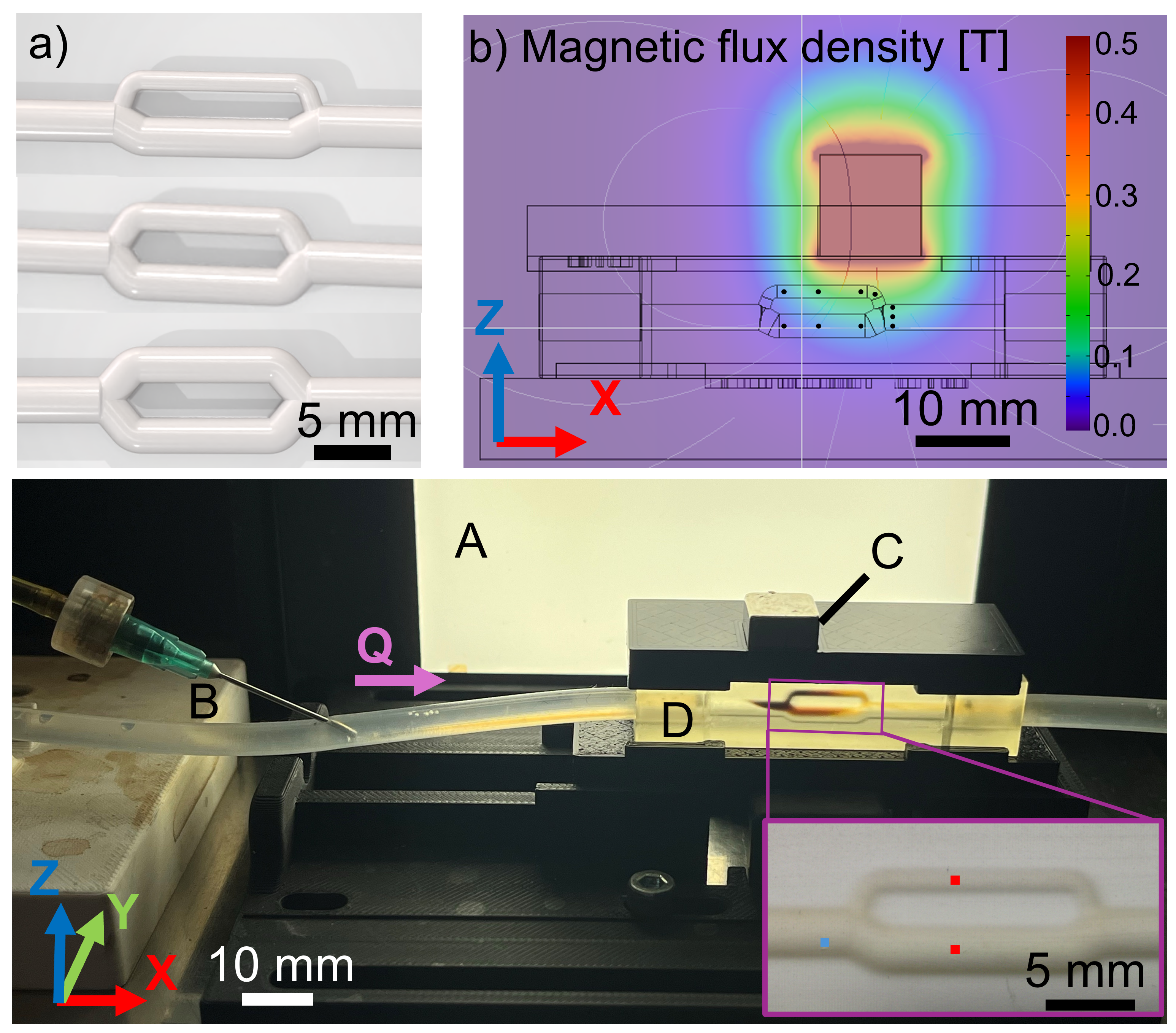}
    \caption{Experimental setup. (a) CAD renderings of the bifurcation geometries with \acp{SF} of 0.25 (top), 0.5625 (middle) and 1 (bottom). (b) Simulated magnetic flux density in the branching plane; black dots mark the evaluation points. (c) Test bench with backlight (A), injection cannula (B), permanent magnet (C) and specimen (D); Q indicates the flow direction. The inset shows the \acp{ROI} in the daughter branches (red) and the reference \ac{ROI} (blue).}
    \label{fig:exp_setup}
\end{figure}
\subsection{Geometry derivation}

The investigated bio-inspired branching structures were designed based on hierarchical scaling relationships derived from fractal analysis of vascular networks \cite{Fleischhauer2026}. The relation between the mother vessel radius $r_0$ and the daughter vessel radii $r_1$ and $r_2$ was first described by Murray as a power law with the radius exponent $\alpha$ according to \eqref{eq_radii_murray}. Based on \cite{Nakamura.2014}, assuming rigid cylindrical vessels with diameters $> \qty{100}{\micro\metre}$, $\alpha = 2.7$ was chosen. Similar scaling relations link the segment length to the vessel radius by a proportionality factor $\Omega$ and an exponent $\kappa$, that were chosen as $\Omega = 10$ and $\kappa = 0.78$ \cite{Nakamura.2014}, while the branching angles were defined according to Tekin et al. \cite{Tekin2016}:
\begin{gather}
    r_0^\alpha = r_1^\alpha + r_2^\alpha \label{eq_radii_murray} \\
    l = \Omega r^\kappa \\
      \cos(\theta_{0}) = \frac{h_{0}^2 - h_{1}^2 - h_{2}^2}{2h_{1}h_{2}} \,, ~~\text{with } h_{i} =  r_i^2 \pi  
\end{gather}
The angle $\theta_0$ denotes the angle between the daughter segments with radius $r_1 ~\mathrm{and } ~ r_2$. The angles $\theta_1 ~\mathrm{and} ~\theta_2$ that represent the angles between the mother vessel and the segments with radius $r_2 ~\mathrm{and } ~ r_1$, respectively, are calculated accordingly \cite{Tekin2016}. Through this definition of the branching angles, asymmetry can be described by the ratio of the daughter vessel dimensions. We therefore define a \acf{SF} relating the cross-sectional areas of the two daughter branches: \(SF = A_2/A_1 = r_2^2/r_1^2\). A \ac{SF} of 1 represents symmetric branching with equal branch sizes, while 0.25 corresponds to a second branch with half the diameter of the first branch. Based on the established scaling laws, all geometric variations were derived from this parameter, and three branched structures with \acp{SF} of $0.5^2$, $0.75^2$ and $1^2$ were investigated.

\subsection{Experimental Setup} \label{sec:exp_setup}
Particle propagation experiments were conducted on a test bench modified from \cite{Fleischhauer2026}. An upstream peristaltic pump provided a background flow at 2.5, 5 and $\qty{10}{\milli\litre\per\minute}$ of distilled water with a density of $\rho_\text{w}=\qty{998}{\kg\per\m^3}$ and a viscosity of $\mu_\text{w}=\qty{1}{\milli\Pa\s}$. The specimen was connected to the periphery by a silicone tube with an inner diameter of $\qty{2.5}{\milli\metre}$. The \acp{SPION} were injected through a $\qty{0.8}{\milli\metre}$ cannula placed $\qty{50}{\milli\metre}$ upstream of the bifurcation with its tip resting on the channel bottom, applying a single pulse of $V_\text{inj}=\qty{0.1}{\milli\litre}$ at a fixed rate of $\qty{10}{\milli\litre\per\minute}$. Particle propagation was recorded by light transmission according to \cite{Fleischhauer2026} using a camera of type Basler BA2-a2A5320-23ucB and a white diffuse LED backlight (see \ref{fig:exp_setup}c). The branching plane was aligned with the x-z-plane, with the entering channel aligned in x-direction, perpendicular to gravity (z-direction), and the smaller-diameter branch pointing upwards. An enclosure ensured constant optical conditions between measurements.
For magnetic actuation, a cubic NdFeB permanent magnet of grade N42 with dimensions \(10 \times 10 \times 10 ~\mathrm{mm}\) was placed above the specimen, centered on the geometric center of the bifurcation. The magnet thus targeted the upper branch, where particle transport is disfavored by stratification and, with increasing asymmetry lower flow fractions. 
The upper branch was exposed to a flux density of about \qty{150}{\milli\tesla} and a gradient of \qty{46}{\tesla\per\metre}, the lower branch to \qty{75}{\milli\tesla} and \qty{18}{\tesla\per\metre}. As a measure of magnetophoretic actuation, \(B|\nabla B|\) yielded values of 6.8 and \qty{1.3}{\tesla\squared\per\metre} for the upper and lower branch, respectively.

\subsection{Particle Properties} \label{sec_particle_properties}
The ferrofluid contains \acp{SPION} with properties similar to previous works \cite{Xiao2026a,Fleischhauer2026}, consisting of a magnetic iron core enclosed by a lauric-acid coating. The particles were synthesized by co-precipitation and purified as described in \cite{Fleischhauer2026}, yielding a master suspension with an iron concentration of $\qty{15.67}{\milli\gram\per\milli\litre}$. Dynamic light scattering yielded a mean hydrodynamic diameter of $D_\text{p}= \qty{40.45}{\nano\metre}$ with a polydispersity index of 0.218. For the experiments, the master suspension was diluted with distilled water to $\qty{10}{\milli\gram\per\milli\litre}$.

\subsection{Image Analysis} \label{sec:image}
Following the procedure described in \cite{Fleischhauer2026}, for each frame, the transmitted-light signal was extracted and quantified by the mean and standard deviation of the pixel intensities within \acp{ROI} of \(10 \times 10~\mathrm{px}^2\), where \(10~\mathrm{px} \approx 100~\mathrm{\mu m}\), positioned at the centers of the daughter channels \qty{580}{px} downstream of the geometric center of the bifurcation. An additional reference \ac{ROI} at the bottom of the feeding channel, \qty{180}{px} upstream of the bifurcation, provided a common timing reference.
Each \ac{ROI} signal was normalized to the mean intensity of the first second of the measurement, when the channel was filled with water, yielding the normalized transmission \(T_n\). According to the Beer--Lambert law, \(T_n\) is related to the local particle concentration \(c_p\) by the optical path length \(d_t\), corresponding to the local channel diameter, and the extinction coefficient \(\beta\):
\begin{align}
    -\ln(T_n) = c_p d_t \beta .
    \label{eq:lambert-beer}
\end{align}
Reference measurements yielded \(\beta = 0.409 \pm 0.006~\mathrm{mL\,mg^{-1}\,mm^{-1}}\) with \(R^2 = 0.983\). For optical densities corresponding to \(c_p d_t > 2\), the calibration deviated from the linear Beer--Lambert relation due to saturation and scattering effects. Concentrations in this regime are therefore reported as apparent concentrations and used only for the assessment of relative trends.

\section{Simulation Setup of the Single Branched Structure} \label{sec:simulation_setup}
In this work, two different simulation approaches are considered, based on prior work. The concentration distribution in the single branched structure was investigated by a \ac{LPS} \cite{Wille2025a} and a \ac{VoFS} \cite{Wille2025}.
The three branch geometries were converted to a 3D mesh, which was numerically solved in the \emph{OpenFOAM} software. The magnetic field was simulated in \emph{COMSOL 6.1} and evaluated at the points of interest within the channel structure. 

\subsection{Definition of the Simulation Regime and Accounted Effects}

\paragraph{Flow Regime}
The flow regime is classified according to the Reynolds number $
    \text{Re} = \rho_\text{w} \bar{u}_\text{F,0} D_0 /{\mu_\text{w}}$
with average fluid velocity $\bar{u}_\text{F,0}$ in the inlet channel \cite{Wille2025a}.
The flow in the injection channel is fixed to \qty{10}{\ml\per\min}, when the injection is active, which leads to a Reynolds number of \qty{297}{}.
%
%
In the inlet channel, the flow is largest when the background flow is at \qty{10}{\ml\per\min}, while the injection is also active. In this case, a maximum Reynolds number of \qty{170}{} is retrieved. 
Therefore, it is concluded that the flow regime is well positioned in the laminar regime, which is the typical case for arteries in this size-range \cite{Jamali2019}. 

\paragraph{Diffusion Regime}
In the nano-scale regime, diffusion can lead to an additional distribution of the \acp{SPION}. The strength of diffusion is expressed in the diffusion coefficient \cite{Jamali2019}, calculated as $D = \frac{k_\text{B} T}{3 \pi \mu_\text{w} D_\text{p}} = \SI{1.0e-11}{\square\metre\per\s}$
with Boltzmann constant $k_\text{B}$ and ambient temperature $T=\qty{20}{\degreeCelsius}$. This diffusion coefficient is comparably small and only led to a negligible change in the simulation result.

\paragraph{Sedimentation}
The sedimentation velocity of a single \ac{SPION} can be estimated with Stokes drag law\cite{Xiao2026a} to $u_\text{sed,stokes} = \frac{1}{18}\frac{\rho_\text{p}-\rho_\text{w}}{\mu_\text{w}} g D_\text{p}^2 = \SI{0.94}{\nano\metre\per\s}$ 
with particle density $\rho_\text{p}=\SI{2047}{\kg\per\cubic\metre}$ and gravitational constant $g=\SI{9.81}{\metre\per\square\second}$. This single particle sedimentation is accounted for in the \ac{LPS} (see Sec. \ref{sec_theory_particle_sim}), but did not influence the results noticeably, since it is very minor.

Based on the work of \cite{Xiao2026a}, it is known that ferrofluids often exhibit stronger sedimentation velocities at the initial stock concentration. In this case, it is preferable to view the \ac{SPION} dispersion as a single fluid entity. This is more accurately reflected by the concentration-based simulation approach (see Sec. \ref{sec_theory_vof_sim}) and will be discussed more in the results section.

\begin{figure}
    \centering
    \includegraphics[width=1\linewidth]{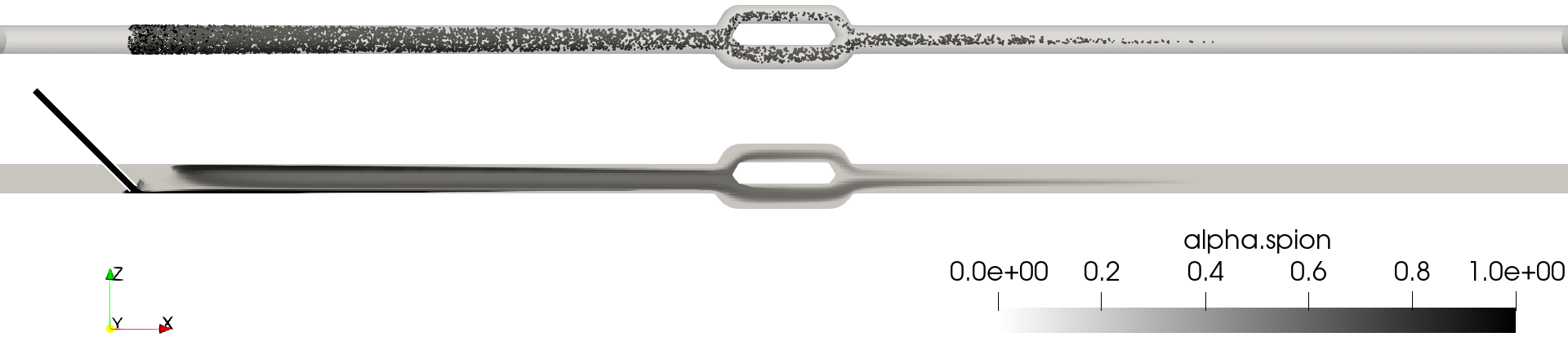}
    \caption{Exemplary \ac{LPS} (top) and \ac{VoFS} (bottom) results showing the predicted ferrofluid volume fraction $\alpha$ in the bifurcation following injection.} 
    \label{fig:sim_methods_example}
\end{figure}

\subsection{Steady Flow and Lagrangian Particle Simulation} \label{sec_theory_particle_sim}
\acl{LPS} is a common approach in the area of \ac{MC} \cite{Jamali2019, Wille2025a}. Particles are treated as single entities and are advected based on a force-balance equation
\begin{equation}
    m_\text{p} \dot{\mathbf{u}}_\text{p}(\mathbf{x},t)=3\pi\mu D_\text{p} \left[ \mathbf{u}_\text{p}(\mathbf{x},t) - \mathbf{u}_\text{F}(\mathbf{x},t)\right] + \mathbf{F}_\text{g}
\end{equation}
with particle mass $m_\text{p}$, particle velocity $\mathbf{u}_\text{p}$, fluid velocity $\mathbf{u}_\text{F}$ and gravitational force $\mathbf{F}_\text{g}$.
In this simulation method, other forces besides gravity are not incorporated. 

This simulation approach is advantageous, as a pre-computed steady velocity field for the background flow can be used to solve for the particle advection.

\paragraph{Simulation Process}
We solved for the steady flow in the different branch geometries with the \emph{SIMPLEC} algorithm and afterwards obtained the particle positions over time with a subsequent Euler time stepping simulation. In the \ac{LPS} (\emph{icoUncoupledKinematicParcelFoam} solver), we reproduced the experiment timings, by first injecting particles steadily over a period of \qty{0.6}{\s} at a distance of \qty{5}{\centi\metre} before the branching. During this injection period a higher inflow of the sum of background flow $Q_\text{bg}$ and injection flow $Q_\text{inj}$ was employed to achieve realistic velocities. Afterwards the inflow was reduced to only $Q_\text{bg}$. The total injected particle amount was set to $N_\text{particle}=\qty{e6}{}$. The simulation time has been adjusted to let at least \qty{99}{\percent} of particles pass through the branches.

\paragraph{Concentration Detection}
The concentration of ferrofluid was obtained from the simulation results by counting the particles in the \acp{ROI} volume $V_\text{DT}$ over time and multiplying with a volume factor $k=\frac{V_\text{inj}}{N_\text{particle}}$.
\begin{equation}
    V_\text{ferro,DT}(t) = k \sum_i^{N_\text{particle}} 
    \begin{cases}
        1, \qquad x_i \in V_\text{DT} \\ 
        0, \qquad \text{otherwise}
    \end{cases} \,.
\end{equation}
Particle $i$ is only counted, when its location $x_i$ is inside the detection volume $V_\text{DT}$.

\subsection{Volume of Fluid Simulation}\label{sec_theory_vof_sim}
The \acl{VoFS} employs an Eulerian approach and treats the particle dispersion as a separate fluid phase, that can mix with the background fluid (water). 
A density of $\rho_\text{ff}=\qty{1006.2}{\kg\per\m^3}$ and viscosity $\mu_\text{ff}=\qty{1.25}{\milli\Pa\s}$ is calculated for the used iron concentration of $\qty{10}{\milli\gram\per\milli\litre}$. The concentration of ferrofluid $\alpha_i(\mathbf{x},t)$ is represented in each mesh cell $i$ and is defined as the ratio of ferrofluid volume $V_{\text{ferrofluid},i}$ to cell volume $V_{\text{cell},i}$ in \eqref{eq_vof_concentration}.
The spatial and temporal change of concentration is solved with the advection equation in \eqref{eq_vof_advection}.
\begin{align}
    \alpha_i(\mathbf{x},t) &= \frac{V_{\text{ferrofluid},i}}{V_{\text{cell},i}} \label{eq_vof_concentration} \\
    \dfrac{d\alpha}{dt} + \nabla \cdot (\mathbf{U} \alpha)& = \nabla \cdot \left(D \nabla\alpha \right) \label{eq_vof_advection}
\end{align}
where $\mathbf{U}$ is the total velocity field. More theoretical details about this specific approach are given by \textcite{Ghorbani2021}. 

In this simulation approach, diffusion is included and we set the diffusion coefficient to $D=\qty{e-11}{\square\meter\per\s}$. Furthermore, sedimentation behaviour is taken into account, as density differences led to larger sedimentation velocities compared to single-particle sedimentation as discussed in \cite{Xiao2026a}.

\paragraph{Simulation Process}
For the \ac{VoFS}, the solver \emph{twoLiquidMixingFoam} was used.
The simulation solved the first \qty{0.5}{\s} by applying only a steady background of water to the inlet to let the flow settle to a steady value. At \qty{0.5}{\s} the injection flow in the cannula inlet (see Fig. \ref{fig:exp_setup}) was ramped up in \qty{0.01}{\s} to the constant value of \qty{10}{\ml\per\min}. The concentration of ferrofluid at the injection inlet was always kept at \qty{1}{}, which led to the influx of ferrofluid in the simulation domain. After \qty{0.6}{\s}, the injection flow was turned off, again in \qty{0.01}{\s}.

\paragraph{Concentration Detection}
The concentration of ferrofluid obtained from the simulation results was integrated in the single \acp{ROI} to obtain the ferrofluid volume over time
\begin{equation}
    V_\text{ferro,DT}(t) = \int_{V_\text{DT}} \alpha(\mathbf{x},t) dV \,.
\end{equation}

\section{Metrics - Linking Experiment and Simulation}

Based on the resulting data from experiment (see Sec. \ref{sec:exp_setup}) and simulation (see Sec. \ref{sec:simulation_setup}) obtained in the respective detection regions, we calculate multiple metrics in order to compare our large data space more easily.

\paragraph{Peak Time}
The peak time is defined as the time point where the signal $f(t)$ exhibits its global maximum, i.e. $f(t_\text{peak}) \geq f(t)\;\forall t$, with the peak value $f(t_\text{peak})$.

\paragraph{Full-Width-Half-Maximum}
The \ac{FWHM} is defined as the time difference $t_2 - t_1$ between the two time points at which the signal reaches \qty{50}{\percent} of the peak value with $t_1 < t_\text{peak} < t_2$.

\paragraph{Decay Rate}
The decay rate is defined as the average signal decrease after the peak with $r_{\mathrm{decay}}=\frac{f(t_{90})-f(t_{10})}{t_{10}-t_{90}}$, where $t_{90}$ and $t_{10}$ denote the time points after the peak at which the signal $f(t)$ has decreased to \qty{90}{\percent} and \qty{10}{\percent} of the global maximum, respectively. For persistent signals that do not reach $t_{10}$ within the measurement period, the decay rate is evaluated from $t_{90}$ to the end of the measurement $t_{\mathrm{end}}$.

\paragraph{Branch Mass Fraction}
To quantify the particle distribution between the daughter branches independently of the pulse shape, a Stewart--Hamilton-based mass proxy is calculated for each branch $i \in \{\mathrm{U},\mathrm{L}\}$ (upper, lower) as
\begin{equation}
    m_i = Q_i \int_{t_0}^{t_0+\Delta t} c_{p,i}(t)\,\mathrm{d}t \,, \qquad Q_i = q_i \, Q \,,
    \label{eq:mass}
\end{equation}
where $c_{p,i}(t)$ is the local particle concentration according to (\ref{eq:lambert-beer}), $Q$ is the inlet flow rate and $q_i$ is the branch flow fraction obtained from the steady flow simulation. The integration starts at the signal onset of the reference \ac{ROI} $t_0$ and covers a common window of $\Delta t = \qty{24}{\second}$ for all measurements. The particle routing is expressed by the upper-branch mass fraction
\begin{equation}
    \varphi_\mathrm{U} = \frac{m_\mathrm{U}}{m_\mathrm{U} + m_\mathrm{L}} \,.
    \label{eq:phi}
\end{equation}
For a transversely homogeneous particle distribution upstream of the bifurcation $\varphi_\mathrm{U} = q_\mathrm{U}$, deviations from the flow fraction indicate the contribution of transverse concentration gradients to the particle routing. Under magnetic retention, stationary particles contribute to the integral in (\ref{eq:mass}), and $\varphi_\mathrm{U}$ is interpreted as an apparent, residence-time-weighted mass fraction.

\section{Results}
\subsection{Hydrodynamic Transport in Experimental Samples} \label{sec_iv_B}
\begin{figure}
    \centering
    \includegraphics[width=1\linewidth]{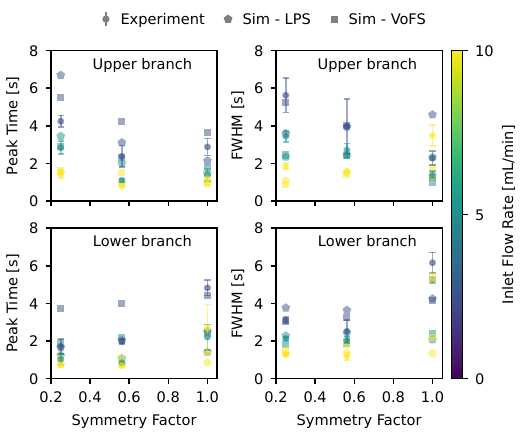}
    \caption{Peak time (left) and \ac{FWHM} (right) in the upper (top) and lower (bottom) daughter branch from experiment, \ac{LPS} and \ac{VoFS}.}
    \label{fig:peak_fwhm}
\end{figure}

In the absence of magnetic actuation, particle propagation was predominantly governed by the underlying flow conditions. In the asymmetric geometries, increasing the inlet flow rate from 2.5 to \qty{10}{\milli\liter\per\minute} consistently reduced both the peak time and the \ac{FWHM} in both daughter branches (Fig.~\ref{fig:peak_fwhm}). The lower-to-upper peak-time ratios of 0.36--0.50 for \(SF=0.25\) and 0.76--0.86 for \(SF=0.56\) reflect the respective hydraulic flow partition. Increasing advection thus reduced both propagation delay and temporal spreading.
The symmetric geometry deviated from this behavior. In the upper branch, the \ac{FWHM} decreased from \qty{2.29}{\second} at \qty{2.5}{\milli\liter\per\minute} to \qty{1.42}{\second} at \qty{5}{\milli\liter\per\minute}, but increased to \qty{3.49}{\second} at \qty{10}{\milli\liter\per\minute}. The lower-branch response remained considerably broader, with \ac{FWHM} values of 6.16, 5.26 and \qty{5.34}{\second}, respectively. In addition, the upper branch exhibited consistently earlier peak times than the lower branch, with peak times of 2.88, 1.40 and \qty{1.19}{\second} compared with 4.83, 2.21 and \qty{2.69}{\second} in the lower branch. However, analysis of the signal onset shows that particles reached the lower branch earlier. Thus, the earlier peak in the upper branch does not indicate faster particle transport, but results from the different temporal signal shape. The upper-branch response is comparatively weak and short and reaches its maximum shortly after particle arrival, whereas the lower branch exhibits an earlier onset followed by a broader and more intense particle pulse whose maximum occurs later. Despite the symmetric flow split, the experimentally observed particle response therefore remains strongly asymmetric, consistent with particle stratification and injection-related concentration gradients.  

\subsection{Experiment - Concentration and Branch-selective Routing} \label{sec:iv_C}

The local peak concentration exhibited pronounced branch-specific differences. In all geometries, the lower branch showed the higher apparent peak concentration, with lower-to-upper ratios of 1.22--1.54 for \(SF=0.25\), 1.11--1.45 for \(SF=0.56\), and 5.72, 3.23 and 2.61 for the symmetric geometry \(SF=1\) at 2.5, 5 and \qty{10}{\milli\liter\per\minute}, respectively. However, peak values are additionally shaped by the branch-specific pulse broadening described in Sec.~\ref{sec_iv_B} and therefore do not directly quantify the particle distribution between the branches. Routing was thus evaluated from the upper-branch mass fraction $\varphi_\mathrm{U}$ according to \eqref{eq:mass} and \eqref{eq:phi} and compared with the flow fraction $q_\mathrm{U}$.

In the asymmetric geometries, $\varphi_\mathrm{U}$ closely followed $q_\mathrm{U}$, with 6.3--\qty{7.5}{\percent} versus 5.7--\qty{6.1}{\percent} for \(SF=0.25\) and 28.7--\qty{32.0}{\percent} versus 28.8--\qty{28.9}{\percent} for \(SF=0.56\). Particle routing was hence governed by the geometric flow partition, and the lower peak concentration in the upper branch resulted from pulse broadening in the slow branch. In contrast, the upper branch of the symmetric geometry received only \qty{8.6}{\percent} and \qty{8.7}{\percent} of the particle mass at 2.5 and \qty{5}{\milli\liter\per\minute}, respectively, despite carrying \qty{50}{\percent} of the flow, which comprises a fraction comparable to that of the small branch of the strongly asymmetric geometry. At \qty{10}{\milli\liter\per\minute}, $\varphi_\mathrm{U}$ increased to \qty{26.0}{\percent}. Thus, despite the balanced hydraulic flow partition, transverse concentration gradients dominated the particle routing at low flow rates. As single-particle sedimentation is negligible on the experimental time scale, these gradients are attributed to collective, density-driven stratification of the injected suspension \cite{Xiao2026a, Fleischhauer2026}. Its decreasing influence with increasing flow rate is consistent with the shorter residence time between injection and bifurcation of approximately \qty{6}{\second} at \qty{2.5}{\milli\liter\per\minute} versus \qty{1.5}{\second} at \qty{10}{\milli\liter\per\minute}.

Two measurement effects have to be considered. Lower-branch signals partly exceeded the linear calibration range, and the total recovered mass proxy increased from 12--\qty{26}{\percent} of the injected mass at \qty{2.5}{\milli\liter\per\minute} to 43--\qty{79}{\percent} at \qty{10}{\milli\liter\per\minute}, suggesting that at low flow rates part of the suspension was transported near the channel bottom outside the central \acp{ROI}. Both effects predominantly reduce the detected lower-branch mass, so that the reported values of $\varphi_\mathrm{U}$ for the symmetric geometry represent upper bounds. 

\subsection{Comparison of Simulative Approaches}
Both numerical approaches predicted the general decrease in propagation time with increasing inlet flow rate but differed in which characteristics of the experimental particle response they captured (Fig.~\ref{fig:peak_fwhm}). In the asymmetric geometries, the \ac{LPS} agreed particularly well with the experimental peak times, with mean absolute deviations across all flow rates and both branches of approximately 0.57 and \qty{0.35}{\second} for \(SF=0.25\) and 0.56, compared with 0.79 and \qty{1.26}{\second} for the \ac{VoFS} approach. The predominantly advective particle transport in these geometries is thus well represented by the particle-based model.

For the symmetric geometry, this relation was reversed, with mean absolute deviations of 1.00 and \qty{0.54}{\second} for the \ac{LPS} and \ac{VoFS}, respectively. While the \ac{LPS} predicted almost identical peak times in both daughter branches, the \ac{VoFS} reproduced the experimentally observed earlier peak in the upper branch for all flow rates. At \qty{10}{\milli\liter\per\minute}, e.g., \ac{VoFS} predicted 1.00 and \qty{1.40}{\second} in the upper and lower branch, compared with experimental values of 1.19 and \qty{2.69}{\second} and particle-based predictions of 0.93 and \qty{0.86}{\second}. \ac{VoFS} therefore captured the direction of the temporal asymmetry, although its magnitude was underestimated. As the dividing streamline of the symmetric bifurcation lies at mid-height, the routing is particularly sensitive to the vertical particle distribution, so reduced depiction of collective sedimentation results in an almost symmetric prediction in the \ac{LPS}. The same applies to the pulse widths. Whereas the \ac{LPS} predicted nearly equal \ac{FWHM} values in both branches, \ac{VoFS} reproduced the broader lower-branch response observed in the experiment (Sec.~\ref{sec_iv_B}), e.g., with 1.25 and \qty{4.15}{\second} in the upper and lower branch at \qty{2.5}{\milli\liter\per\minute}, but underestimates the absolute dispersion.

Each model thus captures a different aspect of the real particle transport. The particle-based description predicts the advective propagation in the asymmetric geometries, while a phase-based description is required to reproduce the stratification-related effects in the symmetric geometry. This is consistent with \cite{Xiao2026a}, where concentrated \ac{SPION} dispersions sedimented as a fluid entity, \ac{SPION} transport thus appears to shift from phase-like behavior at high concentration towards particle-like advection with increasing dilution and flow rate. For the representation of unsteady mixing effects, deviations of the printed branch mesostructure from the ideal geometry, and the slow near-wall transport, a concentration-dependent description combining both regimes is needed.

\subsection{Magnetic Modulation of Particle Propagation}
Magnetic actuation strongly altered the branch-selective particle distribution identified in Sec.~\ref{sec:iv_C}. Since the magnet was placed above the specimen, the magnetophoretic force acts opposite to gravity and thus counteracts the stratification that govern passive routing. For all geometries and flow rates, the apparent upper-branch mass fraction $\varphi_\mathrm{U}$ exceeded the flow fraction $q_\mathrm{U}$, reaching 8.3--\qty{19.1}{\percent} for \(SF=0.25\) and 34.6--\qty{55.2}{\percent} for \(SF=0.56\). In the symmetric geometry, $\varphi_\mathrm{U}$ increased from 8.6, 8.7 and \qty{26.0}{\percent} without magnetic actuation to 55.4, 67.4 and \qty{86.8}{\percent} at 2.5, 5 and \qty{10}{\milli\liter\per\minute}, respectively, exhibiting a reversal of the passive lower-branch preference. In all geometries, the shift towards the upper branch was largest at the highest flow rate, suggesting that particles were also retained in the lower branch at low flow rates, whereas retention was increasingly confined to the upper branch at high flow rates. This branch dependence is consistent with the spatially varying magnetic field, as $B|\nabla B|$ was approximately five times larger in the upper branch. In the strongly asymmetric geometry, this favorable magnetic condition additionally coincided with a substantially lower local flow velocity in the upper branch, with mean velocities of approximately \qty{11.4}{\milli\meter\per\second} and \qty{46.7}{\milli\meter\per\second} in the upper and lower branches at \qty{10}{\milli\liter\per\minute}.

Magnetic actuation also increased the temporal persistence of the particle signal, as evaluated from the post-peak decay rate. For \(SF=0.56\) at \qty{10}{\milli\liter\per\minute}, the decay rate decreased from 0.363 to 0.015~Ext~s\(^{-1}\) in the upper branch and from 0.677 to 0.033~Ext~s\(^{-1}\) in the lower branch. For \(SF=0.25\), the corresponding upper-branch decay rate decreased from 0.080 to 0.014~Ext~s\(^{-1}\). Overall, magnetic actuation modified both the branch-selective particle distribution and the subsequent washout dynamics. The observed response is therefore governed by the local competition between magnetic forcing and advective transport, which is set by the bifurcation geometry through the local flow velocity, superimposing effects of the applied magnetic field and inlet flow rate.

\begin{figure}
    \centering
    \includegraphics[width=1\linewidth]{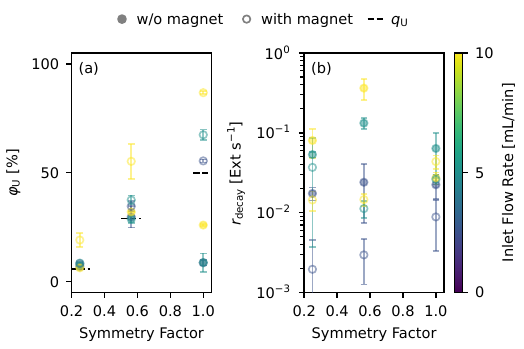}
    \caption{Upper-branch response without and with magnetic actuation. (a) Mass fraction $\varphi_\mathrm{U}$ and flow fraction $q_\mathrm{U}$ (dashed). (b) Post-peak decay rate.} 
    \label{fig:4}
\end{figure}

\section{Conclusion} 
Without magnetic actuation, increasing flow rates reduced propagation delay and dispersion. In the asymmetric geometries, particle routing followed the geometric flow partition, with peak-concentration differences arising from pulse broadening. In the symmetric geometry, stratification dominated, leaving the upper branch only a minor fraction of the particle mass at low flow rates despite carrying half of the flow. The Lagrangian model reproduced the advective propagation, while only the \ac{VoFS} approach captured the stratification-induced pulse deformation. Magnetic actuation counteracted the stratification, reversed the passive branch preference and in most cases reduced the post-peak decay rate, depending on the local balance of magnetic forcing and advection. From a molecular communication perspective, propagation delay, signal dispersion, branch selectivity and particle retention at vascular bifurcations are thus jointly determined by geometry-induced flow partition, collective volume effects and magnetic actuation. Channel models for magnetic targeting should therefore explicitly account for vessel geometry and stratification, using mass-based in addition to peak-based descriptors of branch selectivity. Models that use idealised assumptions potentially neglect the discussed findings and can thus only partly reproduce our results, i.e. they are only valid in some regimes. This work should motivate the inclusion of such effects in the future model development.

\section*{Author Contributions}
DF/LCPW: conceptualization, investigation, validation, visualization, writing–original draft. DF: Experiment. LCPW: Simulation. SL/DD/JK: supervision. All authors: writing–review and editing.



\printbibliography

@article{Tekin2016,
  title = {Do {{Vascular Networks Branch Optimally}} or {{Randomly}} across {{Spatial Scales}}?},
  author = {Tekin, Elif and Hunt, David and Newberry, Mitchell G. and Savage, Van M.},
  editor = {Mac Gabhann, Feilim},
  year = 2016,
  month = nov,
  journal = {PLOS Computational Biology},
  volume = {12},
  number = {11},
  %pages = {e1005223},
  issn = {1553-7358},
  doi = {10.1371/journal.pcbi.1005223}
}

@article{Fleischhauer2026,
  title = {Generalized Blood Vessel Models for Magnetic Nanoparticle-Based Oncology: Geometric and Microfluidic Properties},
  shorttitle = {Generalized Blood Vessel Models for Magnetic Nanoparticle-Based Oncology},
  author = {Fleischhauer, Daniel and Schlicht, Samuel and Drummer, Dietmar},
  year = 2026,
  month = jan,
  journal = {Sci. Rep.},
  volume = {16},
  number = {1},
  pages = {3701},
  issn = {2045-2322},
  doi = {10.1038/s41598-026-37348-7}
}

@article{Wille2025a,
  title = {From {{Steady}} to {{Pulsatile Flow}} in {{Molecular Communication}}: {{Propagation}} of {{Nanoparticles}} in {{Mid-Sized Arteries}}},
  shorttitle = {From {{Steady}} to {{Pulsatile Flow}} in {{Molecular Communication}}},
  author = {Wille, Luiz C. P. and Pfannenm{\"u}ller, Christof and Kirchner, Jens},
  year = 2025,
  month = dec,
  journal = {IEEE Trans. Mol. Biol. Multi-Scale Commun.},
  volume = {11},
  number = {4},
  pages = {531--536},
  issn = {2372-2061, 2332-7804},
  doi = {10.1109/TMBMC.2025.3608558}
}

@article{Jamali2019,
  title = {Channel {{Modeling}} for {{Diffusive Molecular Communication}}---{{A Tutorial Review}}},
  author = {Jamali, Vahid and Ahmadzadeh, Arman and Wicke, Wayan and Noel, Adam and Schober, Robert},
  year = 2019,
  month = jul,
  journal = {Proceedings of the IEEE},
  volume = {107},
  number = {7},
  pages = {1256--1301},
  issn = {0018-9219, 1558-2256},
  doi = {10.1109/JPROC.2019.2919455}
}

@article{Xiao2026a,
  title = {Density {{Matters}}: {{The Effect}} of {{SPION Sedimentation}} on {{Molecular Communication}}},
  shorttitle = {Density {{Matters}}},
  author = {Xiao, Keyu and Wille, Luiz C. P. and Fleischhauer, Daniel and Lotter, Sebastian and Lyer, Stefan and Drummer, Dietmar and Fischer, Georg and Kirchner, Jens},
  year = 2026,
  journal = {IEEE Trans. Mol. Biol. Multi-Scale Commun.},
  volume = {12},
  pages = {875--881},
  issn = {2372-2061, 2332-7804},
  doi = {10.1109/TMBMC.2026.3720566}
}

@inproceedings{Wille2025,
  title = {Effect of the {{Injection Design}} on {{Drug Targeting}} in {{Molecular Communication}}},
  %booktitle = {Proceedings of the 12th {{Annual ACM International Conference}} on {{Nanoscale Computing}} and {{Communication}}},
  booktitle = {Proc. 12th ACM NanoCom},
  author = {Wille, Luiz C. P. and Xiao, Keyu and Kirchner, Jens},
  year = 2025,
  month = oct,
  pages = {179--180},
  publisher = {ACM},
  %address = {University of Electronic Science and Technology of China Chengdu China},
  doi = {10.1145/3760544.3765646}
}

@article{Ghorbani2021,
  title = {Computational Fluid Dynamics Simulation of Buoyant Mixing of Miscible Fluids in a Tilted Tube},
  author = {Ghorbani, M and Giljarhus, K E T and Skadsem, H J and Time, R W},
  year = 2021,
  month = nov,
  journal = {IOP Conference Series: Materials Science and Engineering},
  volume = {1201},
  number = {1},
  pages = {012021},
  issn = {1757-8981, 1757-899X},
  doi = {10.1088/1757-899X/1201/1/012021}
}

@article{Amili.2019,
 author = {Amili, Omid and Golzarian, Jafar and Coletti, Filippo},
 year = {2019},
 title = {In Vitro Study of Particle Transport in Successively Bifurcating Vessels},
 pages = {2271--2283},
 volume = {47},
 number = {11},
 issn = {0090-6964},
 journal = {Annals of biomedical engineering},
 doi = {10.1007/s10439-019-02293-2}
}

@article{Nakamura.2014,
 abstract = {It was analyzed in normal physiological arteries whether the least energy principle would suffice to account for the radius exponent x. The mammalian arterial system was modeled as two types, the elastic or the rigid, to which Bernoulli's and Hagen-Poiseuille's equations were applied, respectively. We minimized the total energy function E, which was defined as the sum of kinetic, pressure, metabolic and thermal energies, and loss of each per unit time in a single artery transporting viscous incompressible blood. Assuming a scaling exponent \textgreek{a} between the vessel radius (r) and length (l) to be 1.0, x resulted in 2.33 in the elastic model. The rigid model provided a continuously changing x from 2.33 to 3.0, which corresponded to Uylings' and Murray's theories, respectively, through a function combining Reynolds number with a proportional coefficient of the l - r relationship. These results were expanded to an asymmetric arterial fractal tree with the blood flow preservation rule. While x in the optimal elastic model accounted for around 2.3 in proximal systemic (r {\textgreater}1 mm) and whole pulmonary arteries (r $\geq$0.004 mm), optimal x in the rigid model explained 2.7 in elastic-muscular (0.1 {\textless} r $\leq$1 mm) and 3.0 in peripheral resistive systemic arteries (0.004 $\leq$ r $\leq$0.1 mm), in agreement with data obtained from angiographic, cast-morphometric, and in vivo experimental studies in the literature. The least energy principle on the total energy basis provides an alternate concept of optimality relating to mammalian arterial fractal dimensions under \textgreek{a} = 1.0.},
 author = {Nakamura, Yoshihiro and Awa, Shoichi},
 year = {2014},
 title = {Radius exponent in elastic and rigid arterial models optimized by the least energy principle},
 %pages = {e00236},
 volume = {2},
 number = {2},
 issn = {2051-817X},
 journal = {Physiological reports},
 doi = {10.1002/phy2.236}
}

@article{Tan.2013,
 author = {Tan, Jifu and Shah, Samar and Thomas, Antony and Ou-Yang, H. Daniel and Liu, Yaling},
 year = {2013},
 title = {The influence of size, shape and vessel geometry on nanoparticle distribution},
 pages = {77--87},
 volume = {14},
 number = {1-2},
 issn = {1613-4982},
 journal = {Microfluidics and nanofluidics},
 doi = {10.1007/s10404-012-1024-5}
}

@article{Thomas.2014,
 author = {Thomas, Antony and Tan, Jifu and Liu, Yaling},
 year = {2014},
 title = {Characterization of nanoparticle delivery in microcirculation using a microfluidic device},
 pages = {17--27},
 volume = {94},
 journal = {Microvascular research},
 doi = {10.1016/j.mvr.2014.04.008}
}

@article{Wicke.2022,
 author = {Wicke, Wayan and Unterweger, Harald and Kirchner, Jens and Brand, Lukas and Ahmadzadeh, Arman and Ahmed, Doaa and Jamali, Vahid and Alexiou, Christoph and Fischer, Georg and Schober, Robert},
 year = {2022},
 title = {Experimental System for Molecular Communication in Pipe Flow With Magnetic Nanoparticles},
 pages = {56--71},
 volume = {8},
 number = {2},
 journal = {IEEE Trans. Mol. Biol. Multi-Scale Commun.},
 doi = {10.1109/TMBMC.2021.3099399}
}


\end{document}